\documentclass[a4paper]{jpconf}
\usepackage{graphicx}
\usepackage{amsmath}
\usepackage{xspace}
\usepackage{xcolor}
\usepackage{tabularx}
\usepackage{booktabs}

\newcommand{\pT}{\ensuremath{p_\mathrm{T}}\xspace}
\newcommand{\GeV}{\ensuremath{\,\mathrm{GeV}}\xspace}
\newcommand{\pion}{\ensuremath{\mathrm{\pi}}\xspace}
\newcommand{\pionzero}{\ensuremath{\pion^0}\xspace}
\newcommand{\esumhit}{\ensuremath{\Sigma E_\textrm{hit}}\xspace}
\newcommand{\GEANTfour}{{\textsc{Geant4}}\xspace}
\newcommand{\HGCAL}{HGCAL\xspace}
\newcommand{\NVIDIA}{Nvidia\xspace}
\newcommand{\abs}[1]{\ensuremath{\lvert #1 \rvert}}
\newcommand{\abseta}{\ensuremath{\abs{\eta}}\xspace}
\newcommand{\edep}{\ensuremath{E_\text{dep}}\xspace}

\begin{document}
\title{GNN-based end-to-end reconstruction in the CMS Phase 2 High-Granularity Calorimeter}

\author{S Bhattacharya$^1$, N Chernyavskaya$^2$, S Ghosh$^3$, L Gray$^4$, J Kieseler$^2$, T Klijnsma$^4$, K Long$^2$, R Nawaz$^5$, K Pedro$^4$, M Pierini$^2$, G Pradhan$^4$, S R Qasim$^{2,5}$, O Viazlo$^6$, P Zehetner$^2$ (on behalf of the CMS Collaboration)}

\address{$^1$ Northwestern University}
\address{$^2$ CERN}
\address{$^3$ Centre National de la Recherche Scientifique}
\address{$^4$ Fermi National Accelerator Laboratory}
\address{$^5$ Manchester Metropolitan University}
\address{$^6$ Florida State University} 

\ead{thomas.klijnsma@cern.ch}

\begin{abstract}
We present the current stage of research progress towards a one-pass, completely Machine Learning (ML) based imaging calorimeter reconstruction.
The model used is based on Graph Neural Networks (GNNs) and directly analyzes the hits in each \HGCAL endcap.
%
The ML algorithm is trained to predict clusters of hits originating from the same incident particle by labeling the hits with the same cluster index.
We impose simple criteria to assess whether the hits associated as a cluster by the prediction are matched
to those hits resulting from any particular individual incident particles.
%
%
The algorithm is studied by simulating two tau
leptons in each of the two \HGCAL endcaps, where each tau may decay according to its measured standard model branching probabilities.
The simulation includes the material interaction of the tau decay products which may create additional particles incident upon the calorimeter.
Using this varied multiparticle environment we can investigate the application of this reconstruction technique and begin to characterize energy containment and performance.
\end{abstract}

\section{Introduction}

The High-Luminosity Large Hadron Collider (HL-LHC) project poses an exceptional challenge for particle-shower reconstruction.
With 200 proton-proton interactions per bunch crossing, it is unclear whether traditional, i.e. hand-written, clustering algorithms can satisfy computing constraints while exploiting the full physics potential of improved detector technologies.
Performing particle reconstruction with machine-learning algorithms can provide a solution by making full use of hardware acceleration and advanced pattern-recognition techniques.
In this paper, we present a machine learning-based reconstruction algorithm for the CMS Phase 2 High-Granularity Calorimeter (\HGCAL). The algorithm is tested in a multiparticle environment derived from simulated tau lepton decays with no pileup.
This algorithm is a stepping stone towards a reconstruction algorithm that yields the necessary physics performance and performs within computing constraints in the projected 200 PU environment.

\section{The CMS Phase 2 High-Granularity Calorimeter}

%
The \HGCAL detector~\cite{CMS:2017jpq} is a sampling endcap calorimeter, comprising 47%
\footnote{%
The number of layers was reduced to 47 in a recent design revision. The version of \HGCAL we simulated comprised 50 layers.
}
sensor and absorber layers with a total thickness of about 10 hadronic interaction lengths ($\lambda$). The first 26 layers correspond to about 25 radiation lengths or 1.3 $\lambda$, consist of silicon sensors and absorber material, and form the electromagnetic section. The sensors are hexagonal in shape and have thicknesses of 120, 200, or 300$\,\mathrm \mu$m, depending on the expected fluence. The sensor cells have areas of about 0.6 and 1.3$\,\mathrm{cm^2}$, with higher granularity closer to the beam pipe. The following hadronic section includes 12 fine sampling layers. The remaining layers have a larger fraction of absorber material. Also here, the sensor size increases with distance from the beam pipe. In regions of lower expected fluences, and therefore increasing distance to the beam pipe and the interaction point, the silicon sensors are replaced by scintillator tiles, regular in pseudorapidity $\eta$ and azimuthal angle $\phi$ and equipped with silicon photomultipliers.

\section{Dataset and ground truth}
\label{dataset}


The events used to train the GNN model are obtained from the CMS detector simulation using \GEANTfour~\cite{Agostinelli:2002hh,Allison:2016lfl} with enhanced tracking of each incident particle's history.
This results in a concise map of simulation-level energy deposits onto reconstructed energy deposits from \HGCAL.
Since the showers of incident particles may overlap entirely, and therefore cannot be feasibly reconstructed separately, this truth map is processed further to account for the limitations of the detector.
To form the final ground truth for training, simulated particles are merged together if they are not expected to be separable by the reconstruction.
This is done by assessing the overlap with neighboring particles starting from the first hit it leaves in the detector.
Based on parameters that can be tuned, adjacent hits in the same layer are collected and the spatial distribution of these hits is used to estimate a shower radius, taking into account the sensor sizes.
If the circular projections of two showers on the front face of \HGCAL overlap, the corresponding particles are merged.
The median (mean) merging radius is 0.28 (0.78) cm in the zero pileup double-tau dataset used for training and testing the model.
Each entry in the dataset consists of about 20000 simulated detector hits, and each detector hit is 5-dimensional (energy, three spatial coordinates, and time).
Every entry contains a different number of detector hits and particles and all hits are used as input to the algorithm.
%
%
Roughly 95\% of the hits in an entry are detector noise, and roughly 65\% of the sum of energies of hits $\edep$ stems from detector noise.
%

\section{Model architecture}

\begin{figure}[h]
\centering
\includegraphics[width=.5\linewidth]{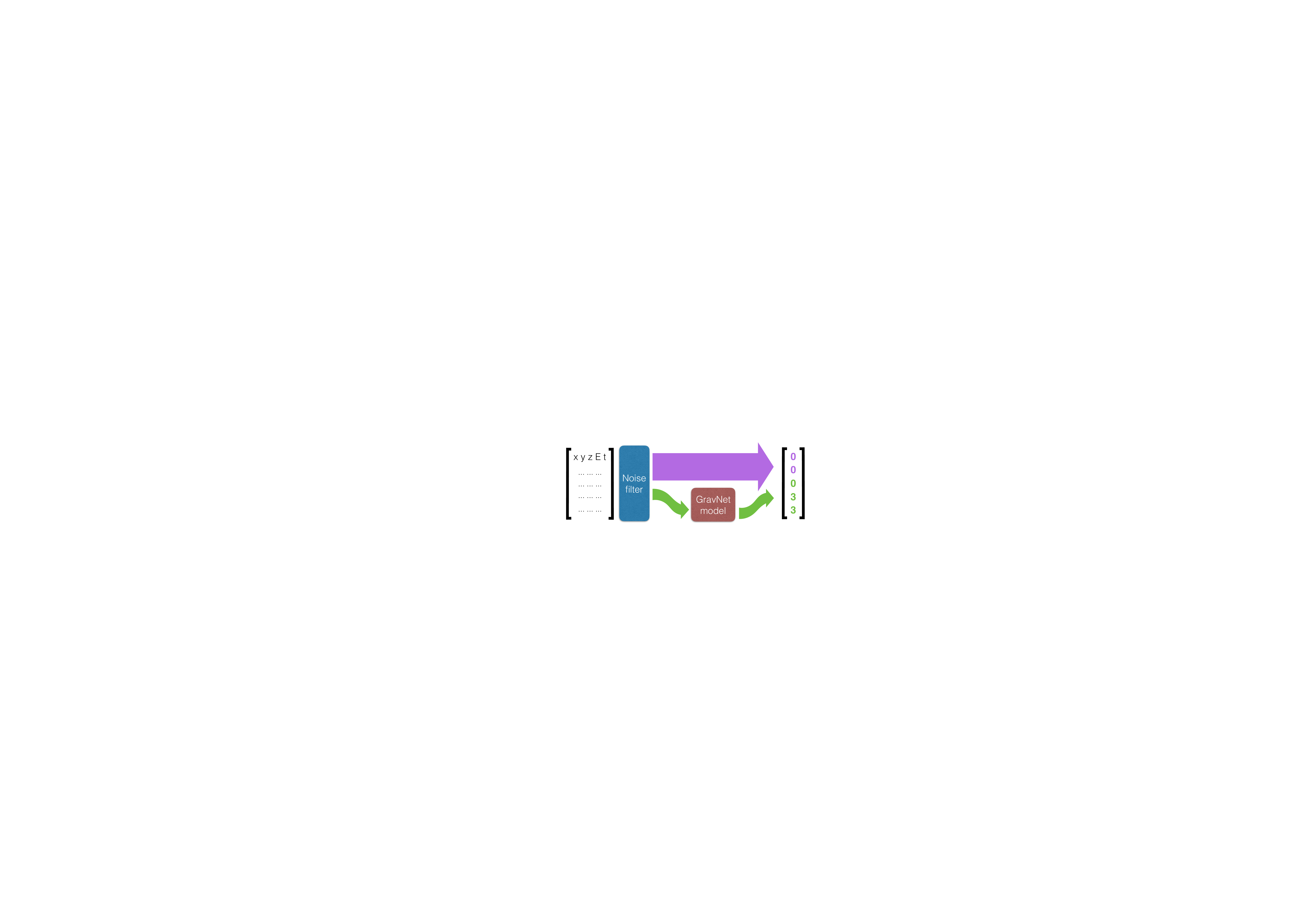}
\caption{\label{model-structure}%
Diagram of the model structure used for the reconstruction algorithm.
The input consists of an $(N \times 5)$ matrix, where $N$ represents the number of hits in the event and 5 is the number of features in a hit.
The input matrix is first passed through a noise filter that removes easily classified noise hits.
The remaining hits are passed to the GravNet model, which performs the reconstruction of hits into particle instances.
Finally, the classifications from the noise filter and the GravNet model are combined into one output matrix.
}
\end{figure}

The reconstruction algorithm is based on a GNN architecture using the ``GravNet''~\cite{Qasim:2021hex} message passing graph convolutional operator and trained using the ``Object Condensation''~\cite{Kieseler:2020wcq} loss function to provide an optimization target that encodes the calorimeter clustering operation.
The model, depicted in Fig.~\ref{model-structure}, consists of two major portions: a simple fully-connected noise filter that masks out most detector noise, and the GravNet model that learns the clustering task on the cleaned data.
These portions are trained together to learn a jointly optimized reconstruction algorithm. The final output of the model in this demonstration is a cluster label
for each hit.

\section{Model training and inference}

The model is trained on 8000 events, each event containing data from both endcaps.
Another 2000 events are used for testing and validation.
The model is trained on an \NVIDIA V100 GPU for 400 epochs, which takes about 8 min per epoch.
The final model is deployed on an \NVIDIA Triton inference server using just-in-time compilation.
Treating the data of one endcap as one inference, we observe a rate of about 32 inferences per second.
By simply overlaying multiple events, we can approximate the event size at 200 pileup.
We then observe a rate of approximately 1 inference per $\mathcal{O}(5)$ seconds.

\section{Results}

Figures~\ref{eventdisplay1} and~\ref{eventdisplay2} show two different viewing angles of an example event display to illustrate the performance of the GNN model.
While close inspection reveals some minor misclassifications, the model provides a generally accurate picture of the original event, even in locally dense environments with overlapping particles.

\newcommand{\eventdisplwidth}{.82\linewidth}

\begin{figure}[h]
\centering
\includegraphics[width=\eventdisplwidth]{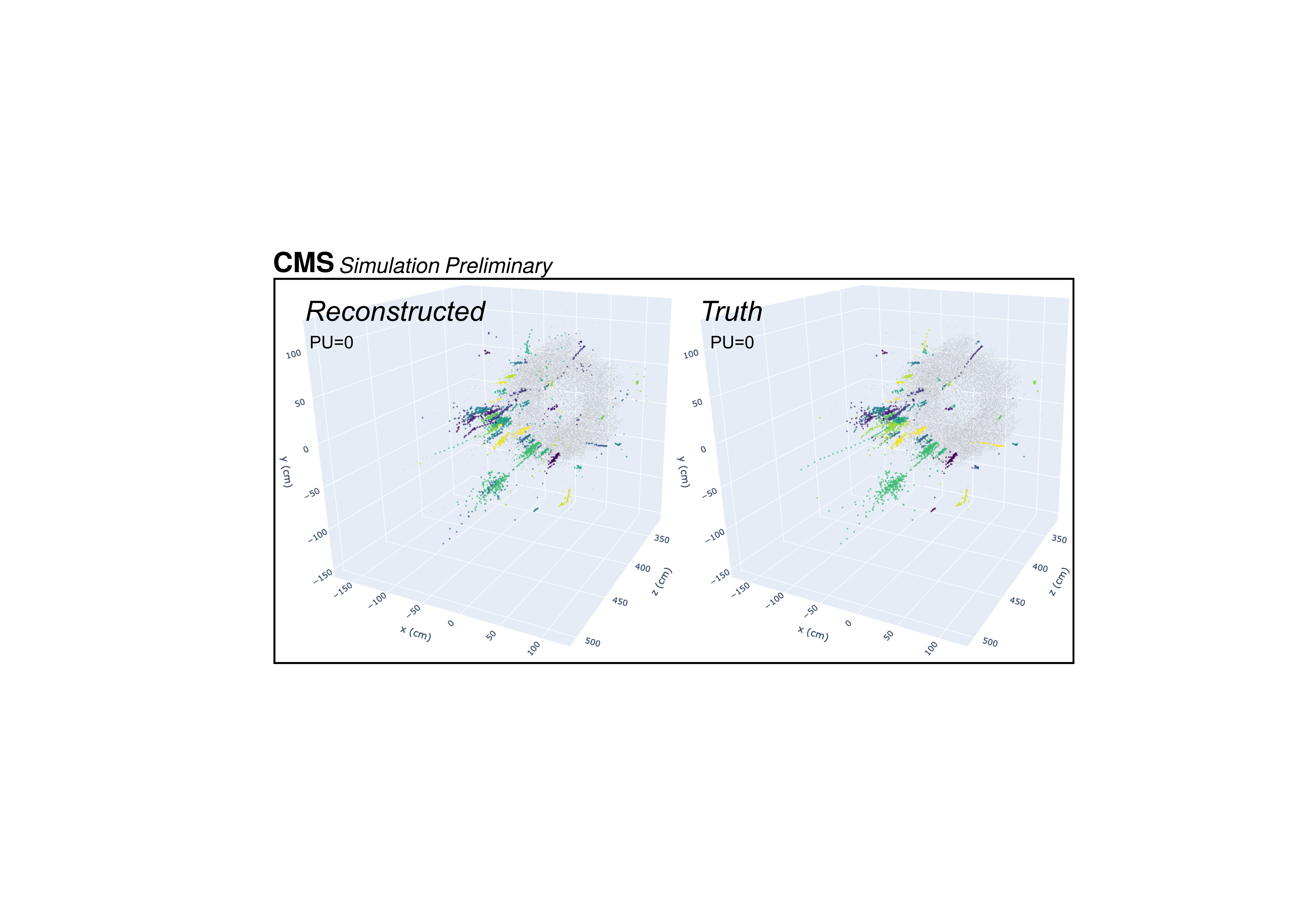}
\caption{\label{eventdisplay1}%
An example in zero pileup of two tau leptons ($\pT = 53.5\GeV$, $\eta = 1.66$; and $\pT = 41.8\GeV$, $\eta = 1.91$) decaying hadronically (to 3\pion, and \pion + \pionzero, respectively) with the decay products entering the \HGCAL detector.
Predicted clusters (left) that overlap with truth showers (right) are given the same color. The grey points are detector noise tagged by the model (left) and at the truth level (right).
The front of the \HGCAL detector is along the back right of the image at $z \sim 300$\,cm, and the showers are evolving out of the page.
The decay products have also interacted with material upstream of \HGCAL, creating sub-showers of various types.
}
\end{figure}

\begin{figure}[h]
\centering
\includegraphics[width=\eventdisplwidth]{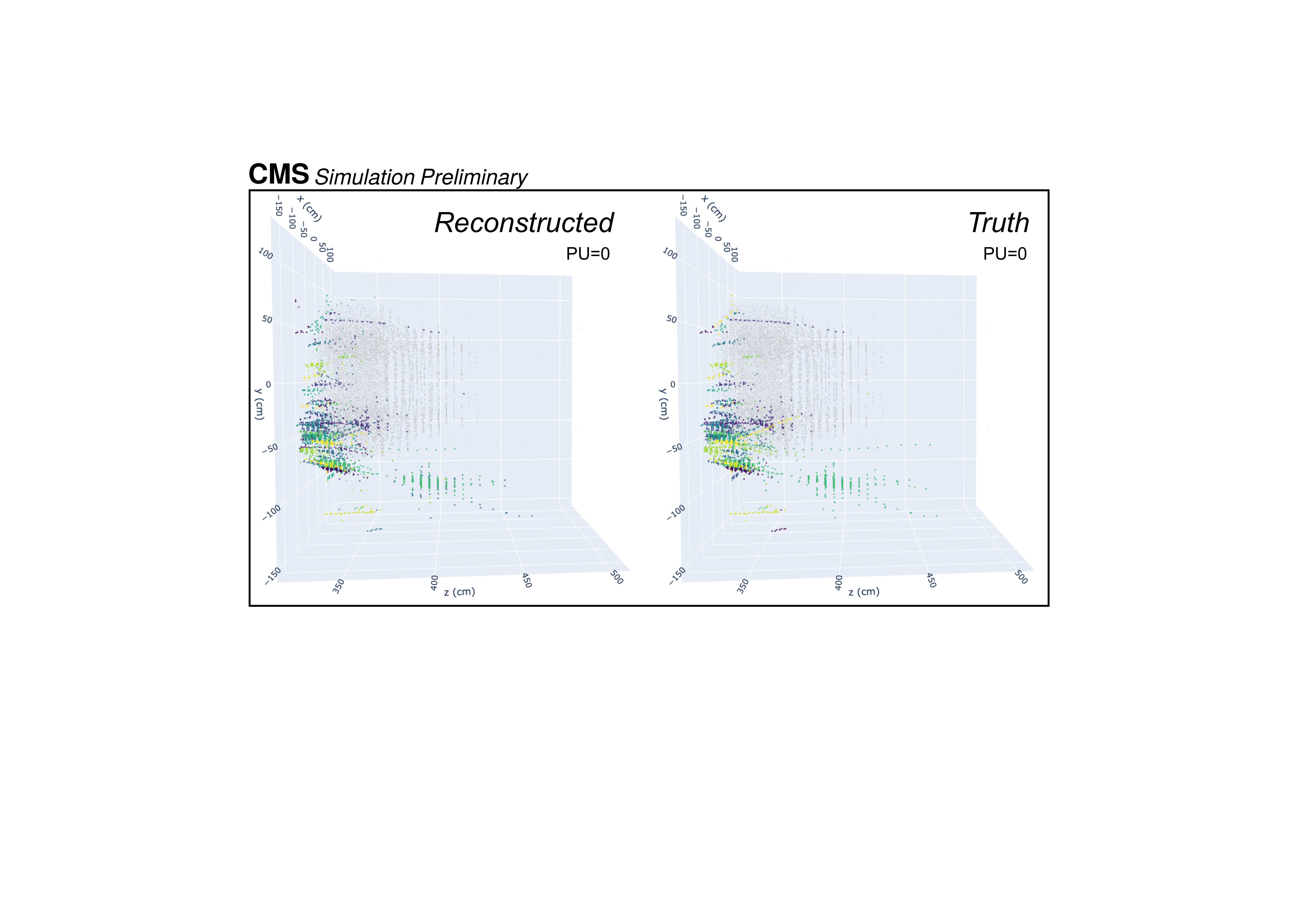}
\caption{\label{eventdisplay2}%
An $r$-$z$ view of the same event as Fig.~\ref{eventdisplay1}, showing the two hadronic tau decays with showers from upstream material interactions.
The front of the \HGCAL detector is to the left at $z \sim 300$\,cm for predicted clusters (left) and truth showers (right).
The longitudinal variation in shower shape for hadronic showers is captured accurately by the prediction, as are the more uniform electromagnetic showers and minimum ionizing particles.
All clusters are predicted in one pass by the GNN reconstruction algorithm.
}
\end{figure}

The reconstruction quality is further evaluated by looking at average quantities over many events.
Table~\ref{table1} provides the average hit energy collection efficiency: the sum of hit energies originating from noise or a real particle, divided by the total energy deposited by the hits.
The top left and bottom right entries are, respectively, the average hit energy fraction that is correctly assigned to originate from noise or from a real particle.
The off-diagonal components are the average energy originating from noise that is reconstructed as a real particle and vice versa. Because these components are small, we conclude that the model has learned to distinguish efficiently between noise and real particles.

The reconstructed clusters are matched to truth showers, as described in Section~\ref{dataset}.
The average fraction of deposited energy that is matched or unmatched is shown in Table~\ref{table2}.
The right column shows the average energy fraction that failed to be matched; the top right entry is the deposited energy from a real particle that was classified as noise, and the bottom right entry is the energy from noise that was classified as a real particle.
With the current matching algorithm parameters, only 0.3\% of the deposited energy from real particles is not matched to a reconstructed particle.

\begin{table}[ht] 
    \renewcommand{\arraystretch}{1.03}
    \setlength\tabcolsep{5pt}
    \centering
    \begin{minipage}[t]{0.48\textwidth}\centering
        \vspace{0pt}
        \begin{tabular}{p{2cm} >{\centering\arraybackslash}p{2.3cm} >{\centering\arraybackslash}p{2.3cm}}
        \toprule
        & $\esumhit$ from reconstructed noise & $\esumhit$ from reconstructed signal \\
        \midrule \\[-11pt]
        $\esumhit$ from truth noise  & 0.637                          & 0.012 \\
        $\esumhit$ from truth signal & 0.004                          & 0.347 \\
        \bottomrule
        \end{tabular} 
        \vspace*{-.1cm}
        \caption{\label{table1}%
            The fractions of the EM-calibrated energy sum of reconstructed hits in the \HGCAL detector tagged by the model as noise or signal.
            %
            %
            The table is normalized to 1.
            The diagonal entries show the correct assignments.
            The amount of signal predicted as noise is minuscule,
            while noise predicted as signal occurs slightly more often.
            }
        \end{minipage}%
    \hfill
    \begin{minipage}[t]{0.48\textwidth}\centering
        \vspace{0pt}
        \begin{tabular}{p{2cm} >{\centering\arraybackslash}p{2.2cm} >{\centering\arraybackslash}p{2.2cm}}
        \toprule
        & Matched & Unmatched \\
        \midrule \\[-11pt]
        $\esumhit$ from truth     & 0.997   & 0.003 \\
        $\esumhit$ from predicted & 0.987   & 0.013 \\
        \bottomrule
        \end{tabular} 
        \vspace*{-.1cm}
        \caption{\label{table2}%
            The fraction of EM-calibrated energy in truth showers that is matched/unmatched with a predicted cluster (top row) and vice versa (bottom row).
            %
            %
            %
            %
            Each row is normalized to 1.
            As can be seen in the top left entry, 99.7\% of the truth energy is matched to a predicted cluster, meaning the GravNet layer treats only 0.3\% of the deposited energy from particles is as noise.
            }
        \end{minipage}
\end{table}

\begin{figure}[h]
\centering
\includegraphics[width=.40\linewidth]{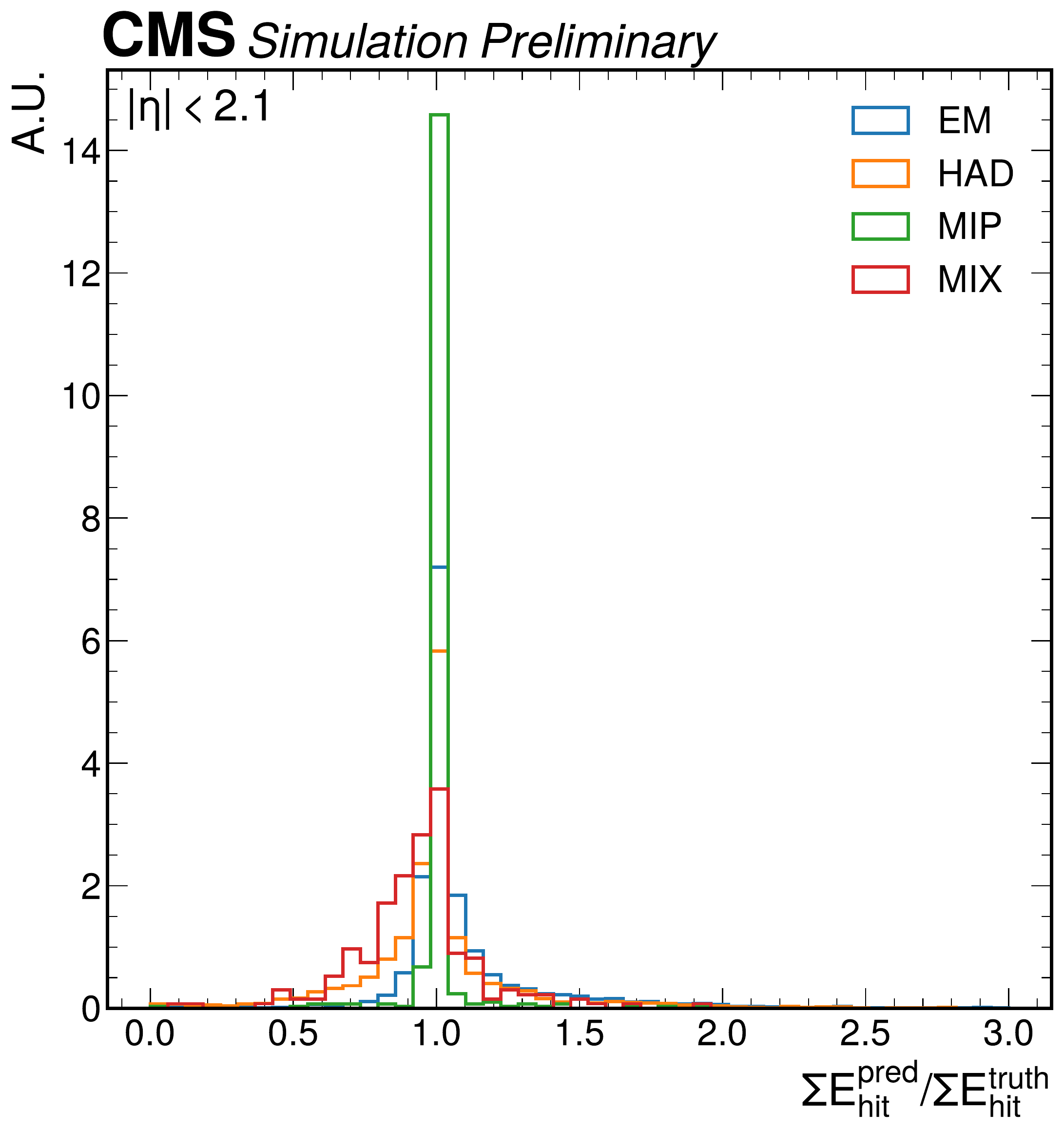}
\includegraphics[width=.40\linewidth]{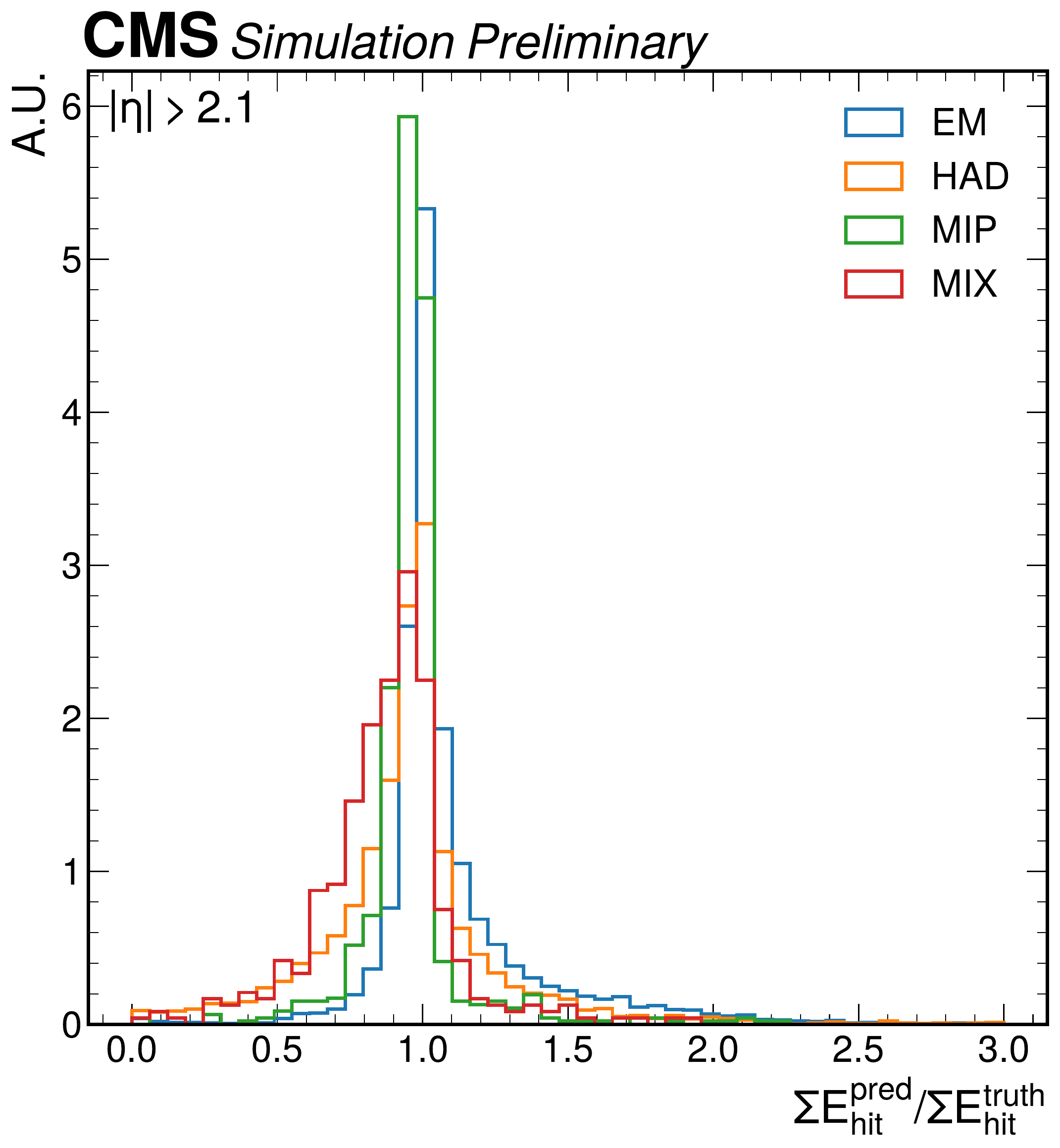}
\vspace*{-.4cm}
\caption{\label{edep-predotruth}%
The ratio of the deposited energy in a reconstructed cluster over the deposited energy in a matched truth shower, for showers in the detector regions with $\abseta<2.1$ (left) and $\abseta>2.1$ (right).
%
%
%
Clusters are partitioned into four categories depending on matched particle species: electromagnetic (EM), hadronic (HAD), minimum ionizing particle (MIP), and truth-matched to multiple particle species (MIX).
}
\end{figure}

In order to evaluate the quality of the individual reconstructed clusters, Fig.~\ref{edep-predotruth} shows deposited energy in a reconstructed cluster over the deposited energy of the matched truth shower, in both the low and high \abseta regions.
The distribution peaks at one as expected, and the distribution is narrower in the low \abseta region.
The distribution is narrowest for the minimum-ionizing particles, followed by the electromagnetic particles and the hadronic particles.
The mixed class of matches, in which truth showers of different classes were matched to a single reconstructed cluster, has the longest tails.

\section{Conclusions and next steps}

We show a promising reconstruction algorithm using graph neural networks for the CMS \HGCAL.
Using the object condensation loss function and a model that composes a noise filter with a state-of-the-art GNN architecture, we have achieved high-performing instance segmentation of particles from tau decays.
Our next step will be to further quantify the physics performance of the network on single, well-understood particles.
Subsequently, we intend to train the model on a dataset with more pileup and more complex physical processes.

\section*{References}
\bibliography{cms}

\providecommand{\newblock}{}
\begin{thebibliography}{1}
\expandafter\ifx\csname url\endcsname\relax
  \def\url#1{{\tt #1}}\fi
\expandafter\ifx\csname urlprefix\endcsname\relax\def\urlprefix{URL }\fi
\providecommand{\eprint}[2][]{\url{#2}}

\bibitem{CMS:2017jpq}
{CMS Collaboration} 2017 {The Phase-2 Upgrade of the CMS Endcap Calorimeter}
  \unskip\space CERN-LHCC-2017-023, CMS-TDR-019

\bibitem{Agostinelli:2002hh}
Agostinelli S {\em et~al.\/} 2003 {\em Nucl. Instrum. Meth. A\/} {\bf 506} 250

\bibitem{Allison:2016lfl}
Allison J {\em et~al.\/} 2016 {\em Nucl. Instrum. Meth. A\/} {\bf 835} 186

\bibitem{Qasim:2021hex}
Qasim S~R, Long K, Kieseler J, Pierini M and Nawaz R 2021 {\em EPJ Web Conf.\/}
  {\bf 251} 03072 (\textit{Preprint} \eprint{2106.01832})

\bibitem{Kieseler:2020wcq}
Kieseler J 2020 {\em Eur. Phys. J. C\/} {\bf 80} 886 (\textit{Preprint}
  \eprint{2002.03605})

\end{thebibliography}
\bibliographystyle{iopart-num}
\end{document}